\documentclass{article}
\usepackage[T1]{fontenc} 
\usepackage[utf8]{inputenc} 
\usepackage{ismir,amsmath,cite,url}
\usepackage{graphicx}
\usepackage{color}
\usepackage{multirow}
\usepackage{comment}
\usepackage{footnote}
\usepackage{booktabs}
\usepackage{float}
\restylefloat{table}

\usepackage{cleveref}
\newcommand{\helena}[1]{\textcolor{red}{#1}}

\usepackage{lineno}

\makeatletter
\newcommand\footnoteref[1]{\protected@xdef\@thefnmark{\ref{#1}}\@footnotemark}
\makeatother

\title{Multiple F0 Estimation in Vocal Ensembles using Convolutional Neural Networks}

\multauthor
{Helena Cuesta$^{1}$ \hspace{1cm} Brian McFee$^{2}$ \hspace{1cm} Emilia G\'omez$^{1,3}$} { 
  $^1$ Music Technology Group, Universitat Pompeu Fabra, Barcelona\\
$^2$ Music and Audio Research Lab \& Center for Data Science, New York University, USA\\
$^3$ European Commission, Joint Research Centre, Seville\\
{\small\tt helena.cuesta@upf.edu}
}

\def\authorname{H. Cuesta, B. McFee, and E. G\'omez}

\begin{document}
\maketitle
\begin{abstract}
This paper addresses the extraction of multiple F$0$ values from polyphonic and \emph{a cappella} vocal performances using convolutional neural networks (CNNs). We address the major challenges of ensemble singing, i.e., all melodic sources are vocals and singers sing in harmony. We build upon an existing architecture to produce a pitch salience function of the input signal, where the harmonic constant-Q transform (HCQT) and its associated phase differentials are used as an input representation. 
The pitch salience function is subsequently thresholded to obtain a multiple F$0$ estimation output. For training, we build a dataset that comprises several multi-track datasets of vocal quartets with F$0$ annotations.
This work proposes and evaluates a set of CNNs for this task in diverse scenarios and data configurations, including recordings with additional reverb. Our models outperform a state-of-the-art method intended for the same music genre when evaluated with an increased F$0$ resolution, as well as a general-purpose method for multi-F$0$ estimation. We conclude with a discussion on future research directions.

\end{abstract}
\section{Introduction}
\label{sec:introduction}
\begin{figure}
 \centerline{
 \includegraphics[width=0.95\columnwidth]{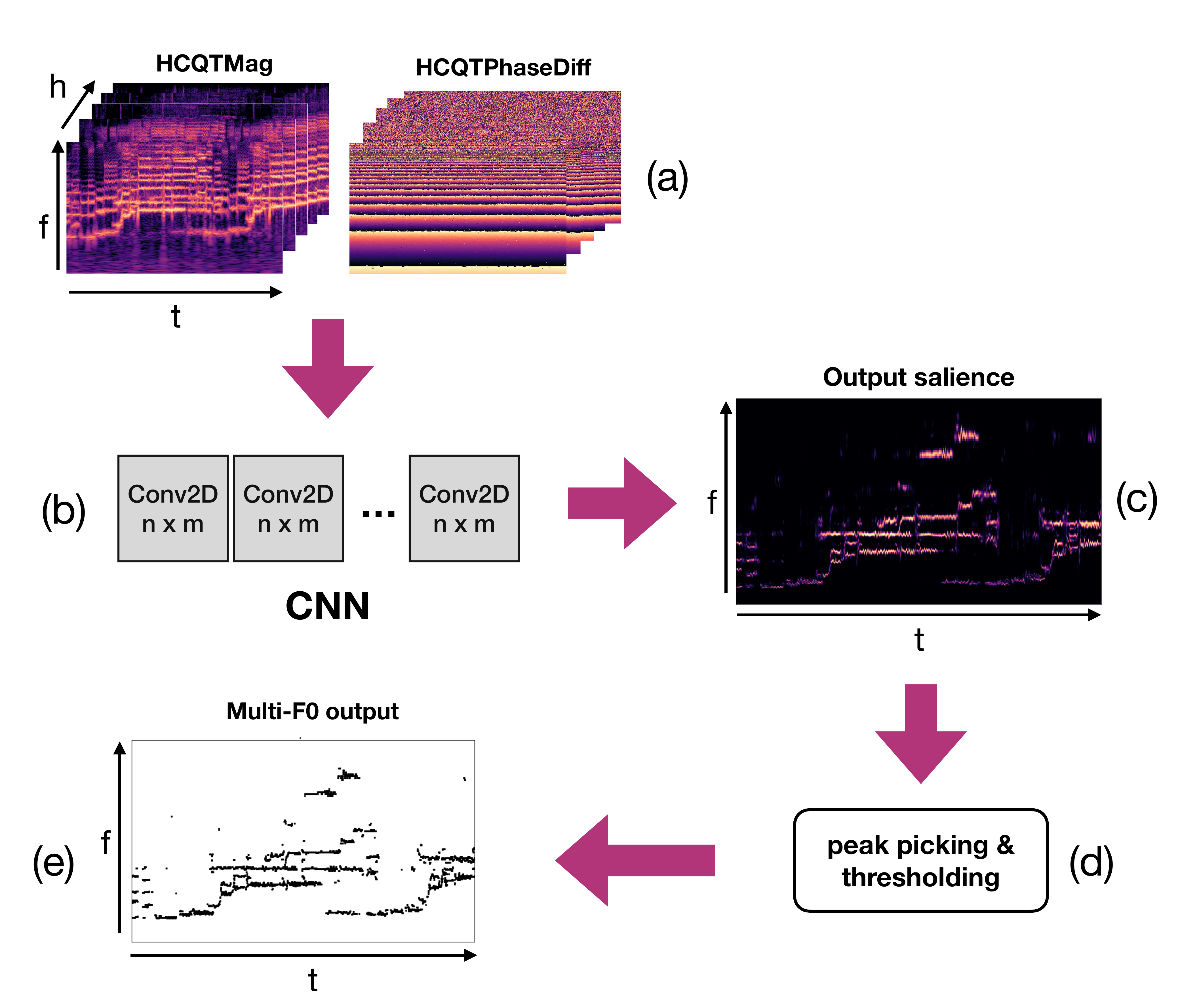}}
 \caption{Overview of the proposed method. \textbf{(a)} Input features. \textbf{(b)} Convolutional architecture diagram. \textbf{(c)} Output salience map. \textbf{(d)} Peak picking and thresholding step to obtain F$0$ values from the salience activation in (c). \textbf{(e)} Multiple F$0$ representation, output of the framework.}
 \label{fig:overview}
\end{figure}
Ensemble singing is a well-established practice across cultures, found in a great diversity of forms, languages, and levels. However, all variants share the social aspect of collective singing, either as a form of entertainment or expressing emotions. In \emph{The Science of the Singing Voice}~\cite{Sundberg87_ScienceSingingVoice}, Sundberg claims that choral singing is one of the most widespread types of singing. In Western classical music, a \emph{choir} is usually a group of singers divided into four sections: soprano, alto, tenor, bass (SATB); however, there exist many other forms of polyphonic singing, involving a diverse number of singers, parts, and vocal ranges. 
One example of such variants is a vocal quartet, where four singers---commonly with distinct vocal ranges---sing in harmony. Vocal quartets usually follow the SATB configuration; therefore, they are different from a \emph{standard} choir in that there is only one singer per section. 
%

Ensemble singing has not been widely studied in the field of Music Information Retrieval (MIR) in the recent years. We find a few early works focused on the acoustic properties of choral singing~\cite{RossingST1986_SopranoChoirSolo,Ternstrom91_PerceptualEvalVoice,Ternstrom02_ChoirAcoustics}, and also a few more recent studies about some expressive characteristics of polyphonic vocal music, such as singer interaction and intonation~\cite{ZadigFL2016_MultitrackChoralInteraction, DevaneyMF12_IntonationSinging_ISMIR, DaiD17_IntonationSATB_ISMIR, CuestaGML18_ChoirIntonation_ICMPC, DaiD19_SingingTogether, WeissSRM19_IntonationLocusIste}, analysis of unison singing~\cite{CuestaGC19_FrameworkMultiF0Choir, ChandnaCG20_AnalysisSynthesisUnison_ISMIR}, or choir source separation~\cite{PetermannCCBGG20_SourceSeparationChoir_ISMIR}.
Most of these studies rely on individual recordings of each voice in the ensemble, which enable the automatic extraction of fundamental frequency (F$0$) contours from each isolated voice. 
The general applicability of these approaches is limited by the fact that vocal groups are rarely recorded with individual microphones.
Another common strategy is using the polyphonic audio recordings along with their associated synchronized scores. However, the process of synchronizing a choral audio recording to a score is not straightforward, and therefore results are not always entirely trustworthy. Manual annotation is also one solution to obtain reliable F$0$ contours, but the process is highly time-consuming and expensive. 

An alternate approach is to analyze a mixed, polyphonic recording to produce multiple F$0$ estimations simultaneously.
This process enables the use of polyphonic singing recordings in the wild, and eliminates the need for separate audio recordings for each singer.
However, multi-F$0$ estimation in polyphonic vocal music has been less often studied, likely due to the complexity and variety of sounds that a singer can produce, the timbre similarity between singers' voices, as well as the scarcity of annotated data of this kind~\cite{SchrammB17_TranscriptionACapellaMultiSingers}.

In this paper, we focus on multiple F$0$ estimation for vocal ensemble audio mixtures. We build upon previous work on using neural networks to obtain an intermediate salience representation suitable for several tasks, including multi-F$0$ estimation~\cite{BittnerMSLB17_DeepSalience_ISMIR}. 
In particular, we experiment with a set of convolutional neural networks (CNN), and combine magnitude and phase information, which is commonly neglected in the literature. We experiment with different fusion strategies and analyze the generalization capabilities of deep learning models in the presence of unison and reverbs.
Figure~\ref{fig:overview} shows a diagram of the proposed method.

Following research reproducibility principles, data generation scripts and models are accessible\footnote{Companion code and models: \url{https://github.com/helenacuesta/multif0-estimation-vocals}}.

%
%
\section{Related Work}\label{sec:related_work}
Multiple F$0$ estimation (also referred to as multi-F$0$ estimation, or multi-pitch estimation) is a sub-task of automatic music transcription (ATM) that consists of detecting multiple concurrent F$0$ values in an audio signal that contains several melodic lines at the same time~\cite{BenetosDGKK13_MusicTranscription_JIIS, BenetosDD18_AutomaticTranscriptionOverview,SchrammB17_TranscriptionACapellaMultiSingers}.
Benetos et~al.~\cite{BenetosDD18_AutomaticTranscriptionOverview} summarize the main challenges of ATM as follows: polyphonic mixtures having multiple simultaneous sources with different pitches, loudness, and timbre properties; sources with overlapping harmonics; and the lack of polyphonic music datasets with reliable ground truth annotations, among others.
They organize ATM approaches into four categories: frame-level (or multi-F$0$ estimation), note-level (or note-tracking), stream-level (or multi-F$0$ streaming), and notation-level.
Our work focuses on the first category.

While monophonic F$0$ estimation is a well-researched topic, with state-of-the-art systems with excellent performances \cite{DeCheveigneK2002_YIN, CamachoH2008_SWIPE, MauchD14_pYIN_ICASSP, KimSLB18_CREPE_ICASSP}, multiple F$0$ estimation is still challenging. Research on this topic is commonly divided into several groups according to the nature of the employed methods. For instance, in~\cite{BenetosDD18_AutomaticTranscriptionOverview} they report four categories: traditional signal processing methods, probabilistic methods, non-negative matrix factorization (NMF) methods, and neural networks.

Klapuri~\cite{Klapuri06_MultiF0_ISMIR} proposed a signal processing based method for multi-F$0$ estimation in polyphonic music. He calculates the salience of F$0$ candidates by summing the amplitudes of its harmonic partials, and then uses an iterative method where at every step an F$0$ is estimated and cancelled from the mixture before moving to the next iteration to estimate the next F$0$. The same author presented in~\cite{Klapuri2008_MultipitchMusicSpeechAuditory} a similar method that incorporates information about human perception by means of an auditory model before the iterative process.

The system presented by Duan et~al.~\cite{DuanPZ10_MultipleF0SpectralPeaks} uses maximum-likelihood approach with the power spectrum as input. Spectral peaks are detected and two separate regions are defined accordingly: the peak region and the non-peak region, using a tolerance of half semitone from the detected peaks.
In the maximum-likelihood process, both sets are treated independently, and the process of detecting F$0$ consists of optimizing a joint function that maximizes the probability of having harmonics that explain the observed peaks and minimizing the probability of having harmonics in the non-peak region. 
The F$0$ estimates are post-processed using neighbouring frames' estimates to produce more stable F$0$ contours.

A recent example of a multiple F$0$ estimation framework that employs neural networks is the system by Bittner et al.~\cite{BittnerMSLB17_DeepSalience_ISMIR}, DeepSalience (DS): a CNN trained to produce a multi-purpose pitch salience representation of the input signal. 
It is designed for multi-instrument pop/rock polyphonic music, and it provides an intermediate representation for MIR tasks such as melody extraction and multi-F$0$ estimation, outperforming state-of-the-art approaches in both cases. Following the premise that a pitch salience function is a suitable representation to extract F$0$ values, also exploited in~\cite{Klapuri2008_MultipitchMusicSpeechAuditory, Klapuri06_MultiF0_ISMIR,RyynanenK08_AutomaticTranscriptionCMJ, SalamonG2012_MELODIA}, this work keeps up with the advancements of deep neural networks to build data-driven salience functions. 
They use the harmonic constant-Q transform (HCQT) as input feature, which is a 3-dimensional array indexed by harmonic index $h$, frequency $f$, and time $t$: $\mathcal{H}[h,f,t]$.
It comprises a set of constant-Q transforms (CQT) stacked together, each of them with its minimum frequency scaled by the harmonic index: $h \cdot f_{min}$. 

While the above methods are well-suited for multiple F$0$ estimation in multi-instrumental music, \emph{a cappella} polyphonic vocal music has several particularities that justify the need for dedicated techniques.
One of the most significant challenges of analyzing vocal ensembles is due to harmonies occurring between distinct, overlapping vocal ranges.
The timbre similarity, strong harmonic relationships, and overlapping frequency ranges hinder the extraction of concurrent F$0$ values in such music signals.

McLeod et~al.~\cite{McLeodSSB17_VocalMusicTranscription_AS} present a system for automatic transcription of polyphonic vocal music, which includes an initial step of estimating multiple F$0$s, and a second step of voice assignment, where each detected F$0$ is assigned to one of the SATB voices. They combine an acoustic model based on the factorization of an input log-frequency spectrogram for the multi-F$0$ estimation 
with a music language model based on hidden Markov models (HMM) for the voice assignment step. 
An earlier version of this method was presented in~\cite{SchrammB17_TranscriptionACapellaMultiSingers}; however, in the latter work the authors include a model integration step where the output of the music language model is further used in the acoustic model to improve the estimation of F$0$s. Their results show that integrating both parts of the system improves the performance of the voice assignment, and also of the multi-F$0$ estimation, since it eliminates many false positives.

Su et al.~\cite{SuCY2016_MultipitchTFC_ISMIR} also address some of the aforementioned issues by proposing an unsupervised method for multi-F$0$ estimation of choir and symphonic music.
Their approach uses time-frequency reassignment techniques such as the synchrosqueezing transform (SST), which aims to better discriminate closely-located spectral components, such as  unisons. 
They use an improved technique called ConceFT, which is based on the idea of multi-taper SST, but was proved to estimate instantaneous frequencies in noisy signals more precisely. These methods measure pitch salience and enhance the stability and localization of the F$0$ features needed for multi-F$0$ estimation.

\section{Dataset}\label{sec:datasets}
\begin{savenotes}
\begin{table}
 \begin{center}
 \resizebox{0.95\columnwidth}{!}{%
 \begin{tabular}{ccc}
  \toprule
  \multirow{2}{*}{\textbf{Dataset}} & \multirow{2}{*}{\textbf{\# of songs}} & \textbf{Duration}  \\ 
  & & \textbf{(hh:mm:ss)} \\ 
  \midrule
  Choral Singing Dataset~\cite{CuestaGML18_ChoirIntonation_ICMPC} & $3$ songs & 00:07:14  \\
  Dagstuhl ChoirSet~\cite{RosenzweigCWSGM20_ChoirSet} & $2$ songs & 00:55:30  \\ 
  ESMUC Choir Dataset & $3$ songs & 00:21:08 \\ 
  Barbershop Quartets\footnote{\label{BSQ}\url{https://www.pgmusic.com/barbershopquartet.htm}} & $22$ songs & 00:42:10 \\ 
  Bach Chorales\footnote{\label{BC}\url{https://www.pgmusic.com/bachchorales.htm}} & $26$ songs & 00:58:20 \\ \bottomrule
 \end{tabular}
 }
\end{center}
 \caption{Overview of the datasets used in this paper. The reported durations refer to the original mixtures before re-mixing stems and data augmentation. Dagstuhl ChoirSet and ESMUC Choir Dataset contain several takes per song.}
 \label{tab:datasets}
\end{table}
\end{savenotes}
%
%
%
The lack of an appropriate and large enough annotated dataset has been a bottleneck in the use of machine learning techniques for multiple F$0$ estimation in ensemble singing. 
We address this difficulty by constructing a dataset that comprises several multi-track datasets of polyphonic singing with F$0$ annotations. 
We created a dataset by aggregating several existing multi-track polyphonic singing datasets. Table~\ref{tab:datasets} shows an overview of the characteristics of each dataset individually. In this section, we describe them in more detail, as well as explain the process of data augmentation.

We use five datasets of similar characteristics. First, the Choral Singing Dataset (CSD)~\cite{CuestaGML18_ChoirIntonation_ICMPC}, a publicly available multi-track dataset of Western choral music. It comprises recordings of three SATB songs performed by a choir of $16$ singers, four per section (4S4A4T4B), and it contains separate audio stems for each singer. Besides, it includes F$0$ annotations for each singer, which are automatically extracted and manually corrected. Similarly, the ESMUC Choir Dataset (ECS) is a proprietary dataset that comprises three songs performed by a choir of $13$ singers (5S3A3T2B); it also includes audio stems for each singer and F$0$ annotations. The third dataset with comparable characteristics is the Dagstuhl ChoirSet (DCS)~\cite{RosenzweigCWSGM20_ChoirSet}: it consists of recordings of two songs performed by a choir of $13$ singers (2S2A4T5B), and two different SATB quartets. This dataset also provides the audio stems and automatically extracted F$0$ annotations. Finally, we also add two commercial datasets: the Bach Chorales (BC)\footnoteref{BC} and the Barbershop quartets (BSQ)\footnoteref{BSQ}. They contain 26 and 22 songs, respectively, performed by vocal quartets---SATB in the first case, and tenor, lead, baritone, bass in the second case---as well as automatically extracted F$0$ annotations. 

We exploit the multi-track nature of all datasets to create artificial mixtures of stems. We use PySox~\cite{BittnerHB16_PySox} to create all the possible combinations of singers, with the constraint of having one singer per part (SATB). In parallel, we also generate the multi-F$0$ annotations by combining the individual F$0$ contours of each singer in the mixture. 

Besides creating the audio mixtures from individual recordings, we include two additional steps to improve generalization. 
First, we augment our dataset by means of pitch-shifting individual voices and re-mixing them. Particularly, we use pitch-shifting at a semitone scale: $-2$ to $+2$ semitones from the original signal. Second, our dataset contains two versions of each audio clip: the original one (obtained by mixing together individual stems), and the same song with reverb. We use the \emph{Great Hall} impulse response (IR) from the Room Impulse Response Dataset in Isophonics~\cite{StewartS10_IsophonicRoomImpulse}, and convolve it with the audio mixtures of our dataset. For both tasks, we use MUDA, a software framework for musical data augmentation~\cite{McFeeHB2016_MUDA}.

The dataset consists of $22910$ audio files of diverse durations, from $10$ seconds to $3$ minutes. We split it into training ($75\%$, $17184$ files), validation ($10\%$, $2291$ files), and test ($15\%$, $3435$ files) subsets.
\section{Proposed Method}\label{sec:method}
\begin{figure}
 \centerline{
 \includegraphics[width=0.9\columnwidth]{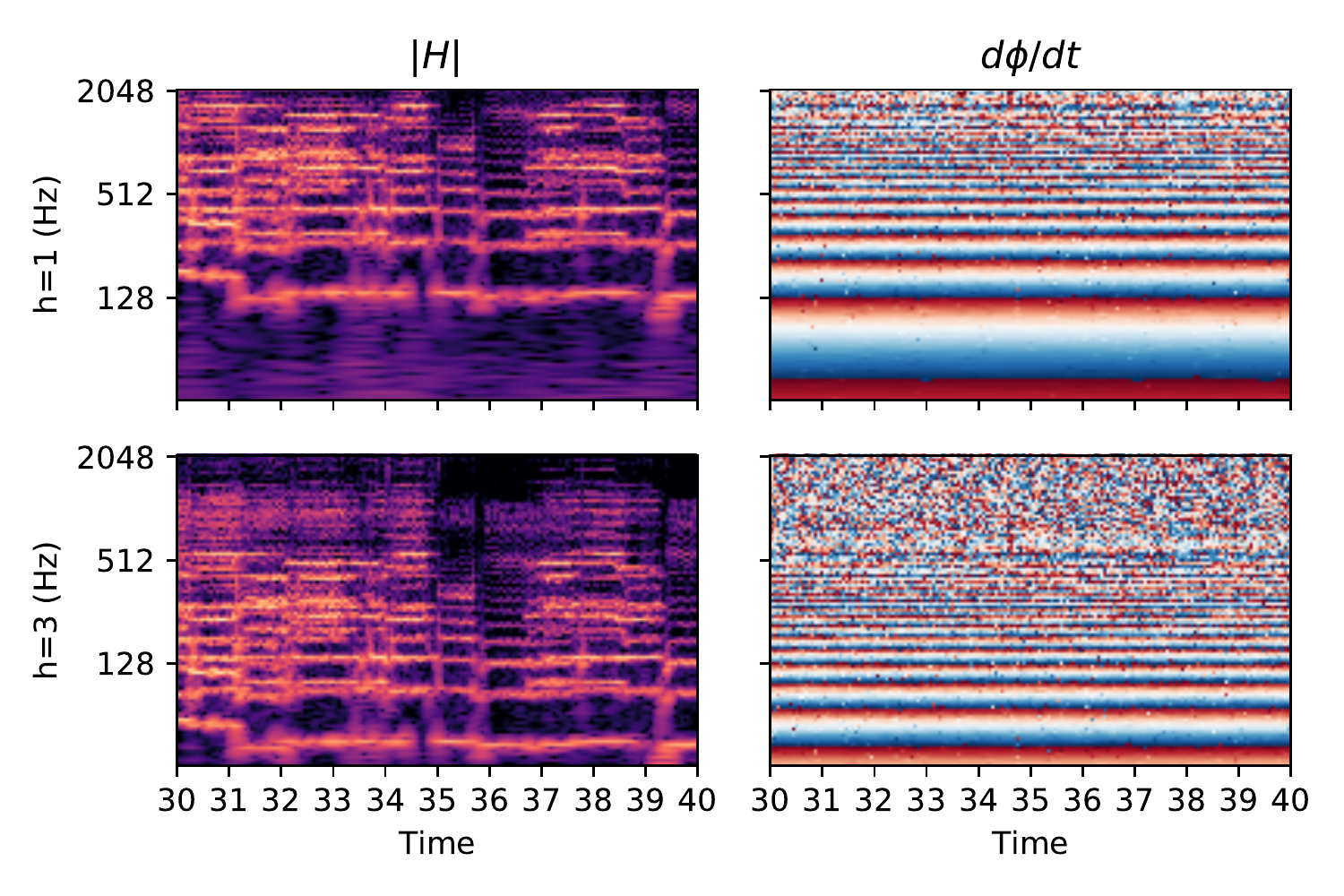}}
 \caption{Input features examples for 10-seconds excerpt of a file from the training set. HCQT magnitude is depicted in the left column, and HCQT phase differentials in the right column, for $h=1$ (top) and $h=3$ (bottom).}
 \label{fig:input_features}
\end{figure}
In this section, we describe the input features, the target representations, the convolutional architectures we design, and the experiments we conduct.
\begin{figure*}
 \centerline{
  \resizebox{0.97\textwidth}{!}{%
  \includegraphics[width=\textwidth]{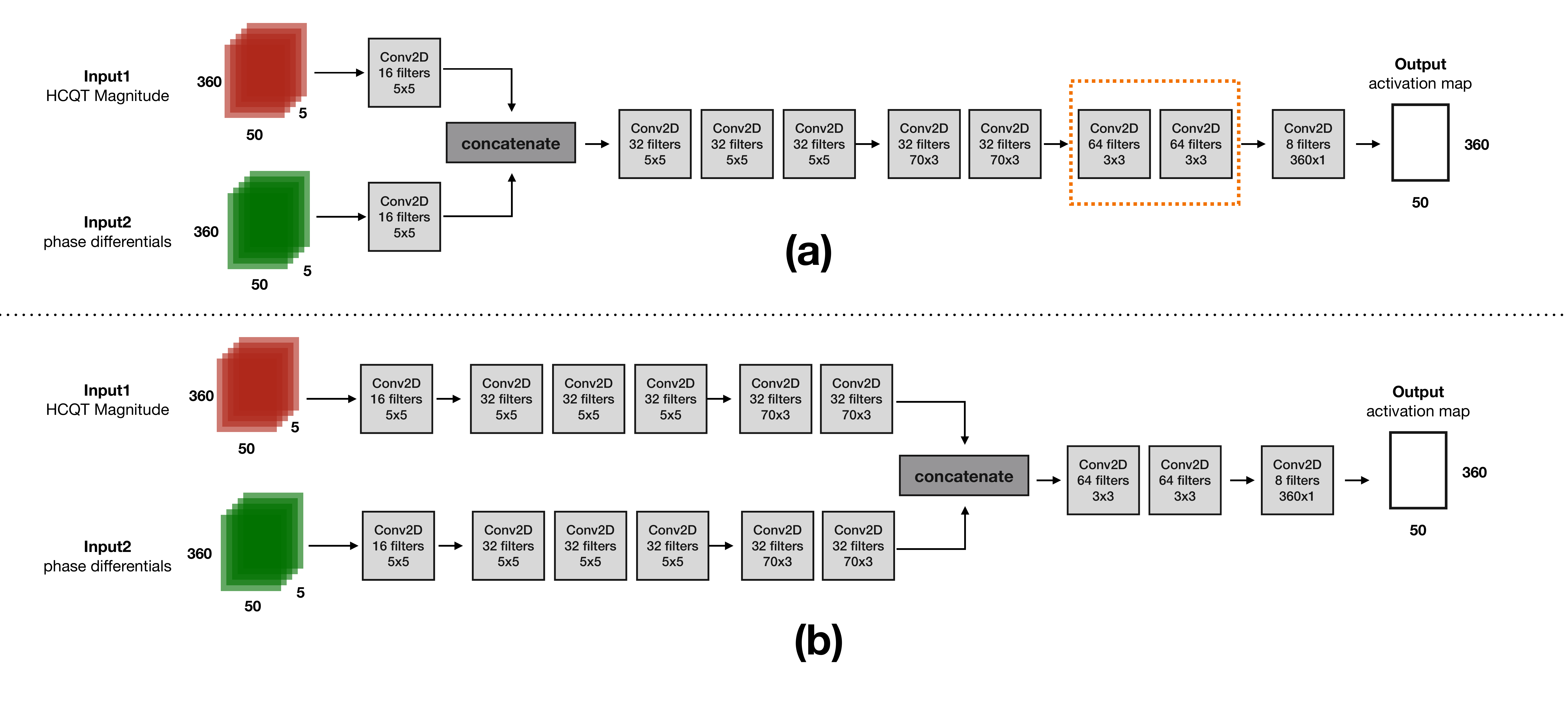}}}
 \caption{Proposed convolutional architectures. \textbf{(a)} Early/Shallow and Early/Deep: the two layers inside the orange dotted rectangle are only part of Early/Deep. \textbf{(b)} Late/Deep: the concatenation of both inputs' contribution happens later in the network. 
 In all networks, each layer is preceded by a batch-normalization step and the output of each layer is passed through a rectified linear unit activation function, except for the last layer, which uses a sigmoid.}
 \label{fig:models}
\end{figure*}
\subsection{Input features}\label{sec:method:inputs}
Our networks have two separate inputs: the HCQT magnitude and the HCQT phase differentials.
The HCQT is a 3-dimensional array $\mathcal{H}[h, t, f]$, indexed by harmonic ($h$), frequency ($f$), and time ($t$). It measures the $h$th harmonic of frequency $f$ at time $t$, where $h=1$ is the fundamental. This representation is based on computing a standard constant-Q transform (CQT) for each harmonic where the minimum frequency ($f_{min}$) is scaled by the harmonic number, $h \cdot f_{min}$. Detailed descriptions of the HCQT are presented in~\cite{BittnerMSLB17_DeepSalience_ISMIR, BittnerMB18_MultitaskLearningF0}.
For the HCQT calculation we use $60$ bins per octave, $20$ cents per bin, $6$ octaves, and a minimum frequency of 32.7 Hz, which corresponds to a C1. We use five harmonics, so that $h \in \{1, 2, 3, 4, 5\}$ to compute the frequencies of the partials. 
Phase information is often discarded from neural network inputs, which commonly use magnitude representations such as the magnitude of the short-time Fourier transform (STFT). However, we also use the associated phase differentials. From signal processing theory we know that the phase differential of a signal contributes to a more precise calculation of the instantaneous frequency ($\omega_{ins}$)~\cite{Boashash92_FreqInsr}:
\begin{equation}
\omega_{ins} = \frac{\delta \varphi(t)}{\delta t}\xrightarrow{}
f_{ins} = \frac{1}{2\pi}\frac{\delta \varphi(t)}{\delta t}
\end{equation}
where $\varphi(t)$ is the phase spectrum of the audio signal.

All audio files are resampled to a sampling rate of $22050$ Hz, and we use a hop size of 256 samples.
An example of the two input features examples is displayed in Figure~\ref{fig:input_features}.

\subsection{Output representation}\label{sec:method:outputs}
The output targets we use to train our networks are time-frequency representations with the same dimensions (2-D) as one of the input channels, i.e., $\mathcal{H}[1]$. We use the ground truth F$0$ annotations (see Section~\ref{sec:datasets}) and assign each F$0$ value to the nearest time-frequency bin in the 2-D representation---which has the same time and frequency resolutions as the input---with a magnitude of $1$. Non-active bins are set to $0$, and we apply Gaussian blur with standard deviation $1$ in the frequency direction to account for possible imprecisions in the predictions. We follow the same procedure as in DeepSalience and set the energy decay from $1$ to $0$ to cover half a semitone in frequency.

\subsection{Models}\label{sec:method:models}
Figure~\ref{fig:models} depicts three convolutional architectures we propose: Early/Shallow, Early/Deep, and Late/Deep.
\subsubsection{Early/Shallow and Early/Deep models}
These models, inspired by DeepSalience, are illustrated in Figure~\ref{fig:models}a. They both consist of a fully convolutional architecture with two separate inputs: one for the HCQT magnitude and a second one for the HCQT phase differentials. Each of these inputs is first sent to a convolutional layer with $16$ ($5\times5$) filters. Then, the outputs of these two layers are concatenated. ($5\times5$) filters cover approximately $1$ semitone in frequency and $50~ms$ in time. After the concatenation, data passes through a set of convolutional layers including two layers with $32$ ($70\times3$) filters, which cover $14$ semitones in frequency and are suitable for capturing harmonic relations within an octave. In the Early/Deep model we add two $64$ ($3\times3$) layers 
before the last layer with $8$ filters that cover all frequency bins.
%
\subsubsection{Late/Deep model}
Late/Deep diagram is displayed in Figure~\ref{fig:models}b, and it follows a similar structure to Early/Shallow and Early/Deep. However, in this case both inputs are handled separately until the layer with ($70\times 3$) filters; then, we concatenate both data streams and add the same layers: two layers with $64$ filters ($3\times 3$), and the last layer with $8$ filters that cover the whole frequency dimension, i.e., $360$ bins.

In all models, batch normalization is applied at the input of every layer, and the outputs are passed through rectified linear units (ReLU), except for the output layer, which uses logistic activation (sigmoid) to map the output of each bin to the range [$0$, $1$]. Using sigmoid at the output enables the interpretation of the activation map as a probability function, where the value between $0$ and $1$ represents the probability that a specific bin belongs to the set of F$0$s present in the input signal.
All models in the experiments described next are trained to minimize binary cross-entropy between the target, $y[t,f]$, and the prediction, $\hat{y}[t,f]$, both of them values in the range [$0$, $1$]:
\begin{equation}
    L(y, \hat{y}) = -y\log(\hat{y}) - (1-y)\log(1-\hat{y})
\end{equation}
We use the Adam optimizer~\cite{KingmaB14_AdamOptimizer} with a learning rate of $0.001$, and train for $100$ epochs with a batch size of $16$ patches of shape ($360, 50$). We perform early stopping when the validation error does not decrease for $25$ epochs.
\subsection{Experimental setup}\label{sec:method:experiments}
\subsubsection{Evaluation metrics}\label{sec:method:metrics}
We evaluate the models using the frame-wise metrics Precision, Recall, F-Score (or F-measure) as they are defined in the MIREX multiple-F$0$ estimation task~\cite{BayED09_MIREX_Multipitch_Metrics} using the \texttt{mir\_eval} library~\cite{RaffelMHSNLE14_mir_eval}.

\subsubsection{Experiment 1: fusion strategy}
In the first experiment we use the whole dataset split into train-validation-test subsets (see Section~\ref{sec:datasets}), to measure the general performance of the three models. With this experiment we study the influence of magnitude and phase information fusion at an early stage of the network, i.e., Early/Shallow-Deep, or later, i.e., Late/Deep. For each model, we use the validation set to optimize the threshold we apply to the peaks extracted from the output salience representation. The optimal threshold is the one that maximizes the average accuracy across the validation set in each case. 
In addition, we train the Late/Deep model without the phase information, i.e., we remove the branch of the network dedicated to the phase. We intend to verify the hypothesis that including the phase as input to the network leads to more precise results.
\subsubsection{Experiment 2: comparative analysis}
In this experiment we evaluate the performance of our best-performing model from Experiment 1 on the BSQ\footnoteref{BSQ}. This is one of the datasets used in~\cite{McLeodSSB17_VocalMusicTranscription_AS, SchrammMSB17_MultipitchVoiceAssignment}, allowing for a direct comparison between their method---also designed for ensemble singing---and the model we propose. These data are part of our original training dataset, but in this experiment we train the model excluding all the BSQ audio files, and then use them for exclusively for evaluation.
\subsubsection{Experiment 3: generalization}
In this last experiment, we aim to explore the effect of unison and reverb. Since vocal ensembles are commonly captured using a room microphone, such recordings usually contain reverb or similar effects, caused by the room acoustics. 
We train our best-performing model excluding all audio files with reverb from the dataset, and then evaluate it with conventional choir recordings from the dataset presented in~\cite{SuCY2016_MultipitchTFC_ISMIR}, which is not part of our working dataset. 
In addition, we evaluate this model on a subset of reverb files from the original test set and compare the performance of this model to the model trained in Experiment 1.
%
%
\section{Results}\label{sec:results}
\begin{figure*}
 \centerline{
 \includegraphics[width=0.95\textwidth]{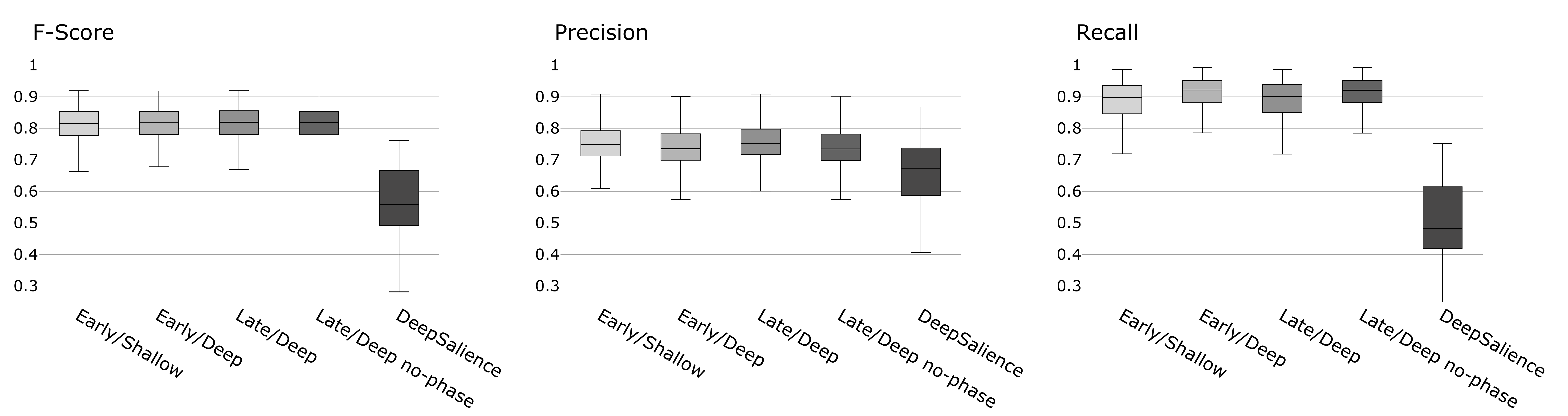}}
 \caption{Evaluation results on the non-pitch-shifted audio files of the test set. We compare our three models to DeepSalience as a baseline, as well as to the Late/Deep network trained without the phase differentials (Late/Deep no-phase).
 Note that outliers are excluded from the plots for an easier visualization.}
 \label{fig:results_exp1}
\end{figure*}
\subsection{Experiment 1: fusion strategy}
\begin{figure}

 \centerline{
 \includegraphics[width=\columnwidth]{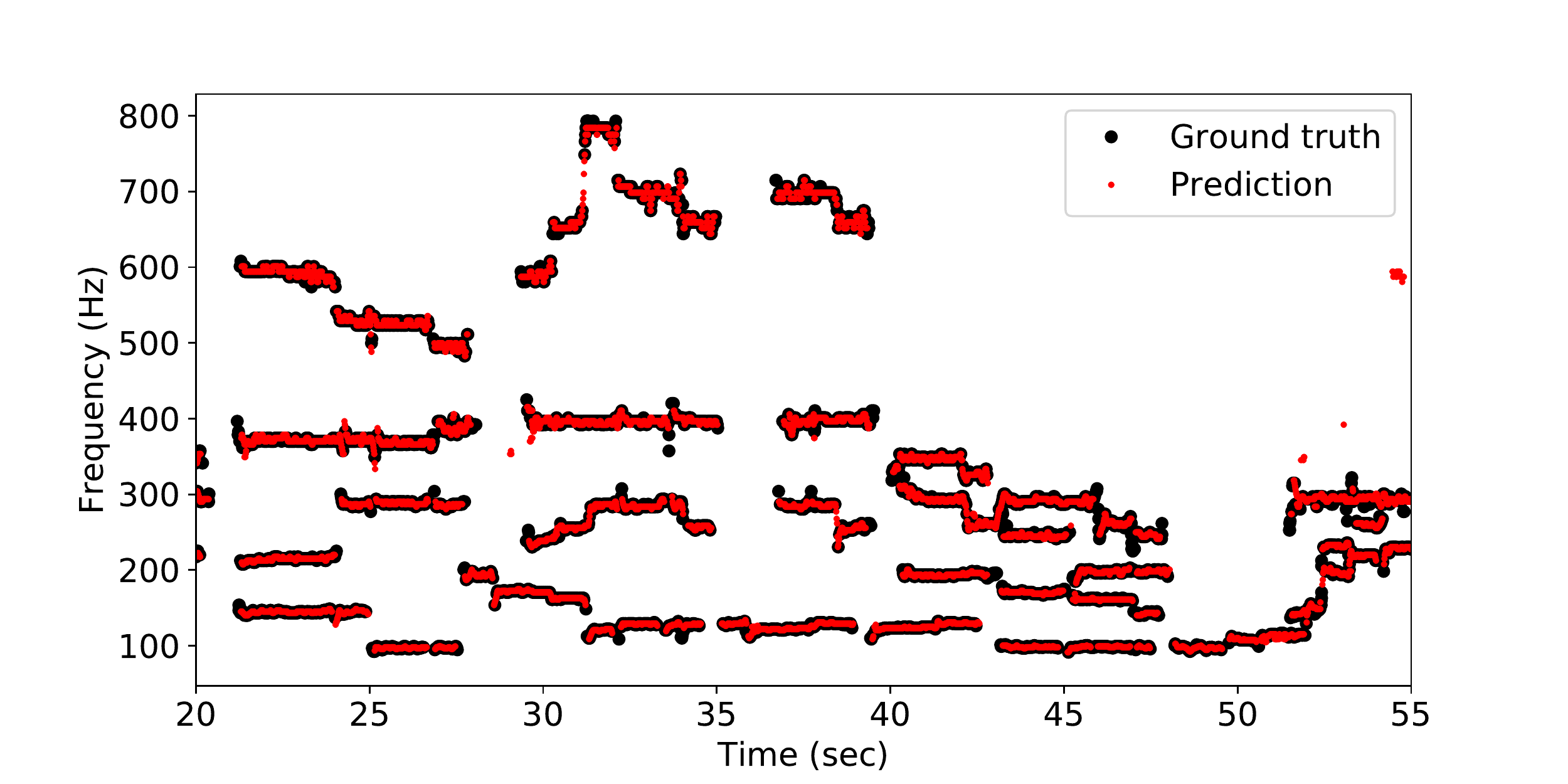}}
 \caption{Example output from Experiment 1, Late/Deep model. Predictions (red) are plotted over the ground truth reference (black).
 }
 \label{fig:output_example}
\end{figure}
Results for Experiment 1 are depicted in Figure~\ref{fig:results_exp1}. While the three models have similar results, Late/Deep is slightly better in terms of F-Score, suggesting that the late fusion of magnitude and phase information is more robust than the early fusion.
Figure~\ref{fig:output_example} shows an excerpt of the multiple F$0$ output (red) together with their associated ground truth reference (black).
We compare these results to DeepSalience using a detection threshold of $0.2$ (optimized beforehand on the evaluation material), which has a lower performance, which we attribute to distribution shift from its training set. 
Additionally, we observe how the Late/Deep without phase information has a similar F-Score but lower precision, showing that including phase differentials as input is helpful for obtaining more accurate results.
Since Late/Deep is the model with the best performance, we use it in the subsequent experiments.

\subsection{Experiment 2: comparative analysis}
Experiment 2 results are summarized in Table~\ref{tab:bsq_comparison} in terms of F-Score, and Precision and Recall (when available). We evaluate the predictions from our model on the BSQ dataset, and compare the results to the ones reported with MSINGERS~\cite{SchrammB17_TranscriptionACapellaMultiSingers}, and VOCAL4-VA, the fully-integrated model from~\cite{McLeodSSB17_VocalMusicTranscription_AS}. 
We use two different pitch tolerances: one semitone ($100$ cents) and $20$ cents. While one semitone resolution is enough for transcription purposes, for analysis such as the ones described in Section~\ref{sec:introduction}, i.e., intonation and singer interaction, more pitch resolution is required.
We observe that our model outperforms both baseline methods with two different pitch tolerances. In the $20$ cents evaluation, the baseline models experience a performance drop ($-17\%$ and $-26\%$), whereas the decrease in our model is much smaller, around $-2\%$ in the F-Score.
This difference presumably resides in the fact that our model uses phase information to refine F$0$ estimates, therefore extracting a more precise value.
\begin{table}[ht]
 \begin{center}
\resizebox{0.98\columnwidth}{!}{%
 \begin{tabular}{ccccccc}
 \toprule
  \multirow{2}{*}{\bfseries Method}& \multicolumn{3}{c}{\bfseries 100 cents} & \multicolumn{3}{c}{\bfseries 20 cents} \\ \cmidrule{2-7} 
    & \bfseries F & \bfseries P & \bfseries R & \bfseries F & \bfseries P & \bfseries R \\ 
    \midrule
  \multirow{2}{*}{MSINGERS~\cite{SchrammB17_TranscriptionACapellaMultiSingers}} & 0.708 & 0.685 & 0.736 & 0.537 & 0.620 & 0.477   \\ 
   &  \textit{(0.06)} & \textit{(0.06)} & \textit{(0.07)} & \textit{(0.07)}& \textit{(0.07)} & \textit{(0.08)} \\ 
   
  \multirow{2}{*}{VOCAL4-VA~\cite{McLeodSSB17_VocalMusicTranscription_AS}} & 0.757 & \multirow{2}{*}{-} & \multirow{2}{*}{-} & \multirow{2}{*}{0.490} & \multirow{2}{*}{-} & \multirow{2}{*}{-} \\ 
  & \textit{(0.06)} & & & & &\\ 
  \multirow{2}{*}{Late/Deep} &\textbf{0.846} & 0.812 & 0.884 & \textbf{0.831} & 0.797 & 0.868 \\ 
  & \textit{(0.03)} & \textit{(0.03)} & \textit{(0.04)} & \textit{(0.03)}& \textit{(0.03)} & \textit{(0.04)} \\ 
  \bottomrule

 \end{tabular}
 }
\end{center}
 \caption{Multi-F$0$ estimation results (F-Score (F), precision (P), and recall (R)) on the Barbershop quartets, for different pitch tolerances.
 Values in parentheses refer to the standard deviation.
 Best scores are highlighted in bold.}
 \label{tab:bsq_comparison}

\end{table}
\subsection{Experiment 3: generalization}
Results from this experiment show that our model outperforms the method in~\cite{SuCY2016_MultipitchTFC_ISMIR} on their choir dataset: when we calculate the average F-Score across the whole dataset, using the threshold optimized on the validation set, we obtain $0.704$, while their best-performing method reaches an average F-Score of $0.653$. Note that this dataset contains short excerpts of commercial choir recordings, with several singers per section and a large reverb effect, which differs from our training material. Therefore, these results suggest that our model is robust to recordings in such context. However, a larger experiment with similar data would be necessary, since this dataset is very small.
The second part of this experiment is the evaluation of a subset of ten reverb files from the test set. In terms of F-Score, and as expected, the model that includes reverb files in the training set (Experiment 1) improves by slightly more than $10\%$ on average with respect to the model that excludes these files (Experiment 3). Therefore, we conclude that the presence of both reverb and dry signals in the training set is beneficial for the performance of a wider range of recording conditions such as reverb. 

\section{Conclusions}\label{sec:conclusions}
In this paper, we proposed a set of novel convolutional architectures for multiple F$0$ estimation in \emph{a cappella} ensemble singing, combining magnitude and phase information.
For training, we created an annotated dataset of polyphonic singing voice by aggregating several existing datasets, and augmented it by means of pitch-shifting and reverberation. 

We conducted several experiments to evaluate different aspects of the detection process. 
We evaluated the overall performance of three models as compared to a deep learning based multi-purpose multiple F$0$ estimation system, and found that our models outperform the baseline when applied on ensemble singing. We also verified that using phase information at the input, together with the magnitude, improves the precision of the F$0$ estimates. 
We compared our best-performing model to one existing approach for multiple F$0$ estimation in vocal ensembles, and demonstrated that it outperforms it with two different F$0$ resolutions ($100$ and $20$ cents).
In addition, we compared our model to an approach specifically designed for choir and symphonic music and found that our model is robust in conditions of unison and high reverb. However, further experiments with a larger amount of data are required to verify these findings.

Although our results are a strong contribution to addressing the limitations of deep learning architectures for vocal music, there are some further steps that would potentially improve the performance of our models. Informal experiments showed that post-processing the output F$0$ contours increases their time continuity; therefore, the overall quality of the output improves. Further steps also include not only estimating the F$0$ values frame-wise, but also assigning each of them to a singer, which is a challenging task if the number of singers is not known a priori.
\section{Acknowledgements}
The authors would like to thank Rodrigo Schramm and Emmanouil Benetos for sharing the BSQ and BC datasets for this research. Helena Cuesta is supported by the FI Predoctoral Grant from AGAUR (Generalitat de Catalunya). This work is partially supported by the European Commission under the TROMPA project (H2020 770376) and MARL-NYU (as part of a two-months research stay).

%

\bibliography{ReferencesMIR}
\end{document}